\documentclass[fleqn,twoside]{article}
\usepackage{espcrc2}
\usepackage{graphicx}

\newcommand{\Eq}[1]{Eq.~(\ref{eq#1})}
\newcommand{\beq}{\begin{equation}}
\newcommand{\eeq}{\end{equation}}

\def\lsim{\;\raisebox{-.6ex}{$\stackrel{<}{\sim}$}\;}
\def\gsim{\;\raisebox{-.6ex}{$\stackrel{>}{\sim}$}\;}

\def\decayarrow{\kern0.2em\hbox{$\raise1.08ex\hbox{\big|}\kern-0.5em
\longrightarrow$}}
\newcommand{\optbar}[1]{\shortstack{{\tiny (\rule[.4ex]{1em}{.1mm})} 
  \\ [-.7ex] $#1$}}

\title {Neutrino Physics: Where Do We Stand, and Where Are We Going?\\
---The Theoretical-Phenomenological Perspective\thanks{Published in the Proceedings of the XXth International Conference on Neutrino Physics and Astrophysics (Neutrino 2002), {\it Nucl. Phys. B (Proc. Suppl.)} 118 (2003) 425 }}

\author{Boris Kayser\\
Fermilab MS106, P.O.Box 500, Batavia IL 60510\\
February 2003}

\begin{document}

\begin{abstract}
The discoveries and open questions in neutrino physics, as reported at Neutrino 2002 and more recently, are reviewed from a theoretical perspective.
\vspace{1pc}
\end{abstract}

\maketitle

\section{THE QUESTIONS}

During the last several years, stunning experimental results have established that neutrinos almost certainly have nonzero masses and mix. This development opens a whole new world for us to explore. What have we already learned about the neutrinos, and what would we like to find out?

Our discussion may be framed in terms of a number of questions:
\begin{itemize}
\item Do neutrinos truly change from one flavor to another?
\item How many neutrino species are there? Do sterile neutrinos exist?
\item What are the masses of the mass eigenstates $\nu_i$?
\item Is each mass eigenstate---
	\begin{itemize} \vspace{-.08in}
	\item A Majorana particle ($\overline{\nu_i} = \nu_i$)
	\end{itemize}
\vspace{-.12in} or
	\begin{itemize} \vspace{-.12in} 
	\item A Dirac particle ($\overline{\nu_i} \neq \nu_i$) ?
	\end{itemize}
\item What are the elements of the leptonic mixing matrix $U$? What mixing angles does this matrix contain?
\item Does $U$ contain CP-violating phases? If so, do these phases lead to detectable CP-violating effects in neutrino oscillation? In neutrinoless double beta decay?
\item Was baryogenesis in the early universe made possible by \emph{leptonic} CP violation?
\item Do the properties of neutrinos and antineutrinos violate CPT invariance?
\item What can neutrinos tell us about astrophysics and cosmology?
\item Can neutrinos serve as probes of extra spatial dimensions beyond the familiar three?
\item What are the electromagnetic properties of neutrinos? What are their dipole moments?
\item How fast do neutrinos decay? Into what do they decay?
\item What is the origin of neutrino flavor physics? Is it new physics at a high mass scale? If so, what is that scale and what physics is found there? Does the see-saw mechanism generate neutrino masses? Do symmetries play a role in neutrino masses and mixing? What is the connection between neutrino flavor physics and quark flavor physics?
\end{itemize}

Let us discuss at least some of these questions. Since this is being written at the end of 2002, we shall try to take post-Neutrino 2002 developments into account.

\section{DO NEUTRINOS TRULY CHANGE FLAVOR?}

By now,  the evidence that neutrinos change from one flavor to another is very strong indeed. At Neutrino 2002, truly striking evidence that the solar neutrinos do this was presented by the Sudbury Neutrino Observatory (SNO) \cite{r2}. As we recall, the nuclear processes that power the sun produce only electron neutrinos, $\nu_e$. 
But the SNO results cleanly establish that the solar neutrino flux arriving at earth includes $\nu_\mu$ and/or $\nu_\tau$. SNO detects high energy solar neutrinos from $^8$B decay using three detection reactions. As summarized in Table \ref{t1}, the observed rates for these reactions measure three different linear combinations of the $\nu_e$ flux, $\phi_e$, and the $\nu_\mu+\nu_\tau$ flux, $\phi_{\mu\tau}$, arriving from the sun. 
\begin{table}[htb]
\caption{The detection reactions employed by SNO, and the fluxes they measure.}
\label{t1}
\begin{tabular}{cc}
\hline \vspace{0.1cm}
\underline{Detection Reaction}    & \underline{Flux Measured} \\
$\nu d \to \nu np$  & $\phi_e + \phantom{0.15\,}\phi_{\mu\tau}$ \\
$\nu e \to \nu e$    & $\phi_e + {0.15\,}\phi_{\mu\tau}$ \\
$\nu d \to epp$        & $\phi_e   \phantom{+ 0.15\,\phi_{\mu\tau}}$ \\
\hline
\end{tabular}
\end{table}
From Table \ref{t1}, it is obvious that from the observed rates, the $\nu_\mu+\nu_\tau$ flux $\phi_{\mu\tau}$ can readily be extracted. SNO finds that  \cite{r2}
\beq
\phi_{\mu\tau} = (3.41\,^{+0.66}_{-0.64} ) \times 10^6 / \mathrm{cm}^2
\mathrm{\,sec}~~,
\label{eq2.1}
\eeq
a result 5.3$\,\sigma$ from zero. When solar $\nu e \to \nu e$ data from the Super-Kamiokande (SK) detector  \cite{r3} are included, the result becomes \cite{r2}
\beq
\phi_{\mu\tau} = (3.45\,^{+0.65}_{-0.62} ) \times 10^6 / \mathrm{cm}^2
\mathrm{\,sec}~~,
\label{eq2.2}
\eeq
5.5$\,\sigma$ from zero. Clearly, solar $\nu_e$ do change into $\nu_\mu$ 
and/or $\nu_\tau$.

Except for higher-order effects that are expected to be negligible, neutrino flavor change does not change the total neutrino flux. It merely redistributes that flux among the different flavors. To be sure, these flavors may include ``sterile'' ones---flavors that do not participate in the normal weak interactions. 
Thus, flavor change can reduce the total \emph{active, detectable} neutrino flux. Let us for the moment assume, however, that there are no sterile flavors---only $\nu_e, \;\nu_\mu$ and $\nu_\tau$. Then the total active solar neutrino flux reaching the earth, $\phi_e + \phi_{\mu\tau}$, should have the value expected if one forgets all about flavor change and just calculates the total rate of neutrino production by the sun. 
The Standard Solar Model (SSM) calculation \cite{r4} of the production rate for $^8$B neutrinos predicts in this way that $\phi_e + \phi_{\mu\tau}$ should have the value \cite{r5}
\beq
\phi_{SSM} = (5.05\,^{+1.01}_{-0.81} ) \times 10^6 / \mathrm{cm}^2
\mathrm{\,sec}~~.
\label{eq2.3}
\eeq
By comparison, SNO finds from its measured $\nu d \to \nu np$ event rate (cf. Table \ref{t1}) that for $^8$B neutrinos reaching the earth \cite{r2}
\beq
\phi_e + \phi_{\mu\tau} = (5.09\,^{+0.64}_{-0.61} ) \times 10^6 / \mathrm{cm}^2
\mathrm{\,sec}~~.
\label{eq2.4}
\eeq
While the uncertainties in these fluxes are obviously not negligible, the agreement between them is quite gratifying. It provides evidence that solar neutrino production is correctly understood, and makes the case that neutrinos change flavor still more compelling.

From the SNO data, $\phi_e/(\phi_e + \phi_{\mu\tau}) \simeq 1/3$. That is, more than half of the $^8$B $\nu_e$ flux created in the solar core changes flavor.

Barring non-Standard-Model flavor-changing neutrino-matter interactions \cite{r6}, neutrino flavor change implies neutrino mass and mixing. Thus, neutrinos almost certainly have masses and mix. This means that the leptons, including the neutrinos, are much like the quarks. In particular, just as there is a unitary matrix $V$ that describes quark mixing, so there is a unitary matrix $U$ that describes lepton mixing \cite{r7}. Complex phases in $U$, as in $V$, can lead to CP violation.

At the time of Neutrino 2002, several candidate mechanisms for the observed solar neutrino flavor change were being considered. The candidates included various versions of the Mikheyev Smirnov Wolfenstein (MSW) effect \cite{r8} within the sun, neutrino oscillation in the vacuum between the sun and the earth, and a number of non-standard scenarios \cite{r6}.  
In response to the SNO measurements of solar fluxes via all three of the reactions in Table \ref{t1}, including for the first time the neutral-current reaction $\nu d \to \nu np$, several analyses of all the solar neutrino data had been performed \cite{r9,r10}. These were extensively discussed at Neutrino 2002 \cite {r10}. 
There was general agreement that the ``standard'' candidate most favored by the data was the Large Mixing Angle (LMA) version of the MSW effect. However, at that time there were other candidates that were not excluded.

Since Neutrino 2002, the physics behind the behavior of solar neutrinos has been greatly clarified by the KamLAND experiment. KamLAND studies the flux of $\overline{\nu_e}$'s from nuclear power reactors that are typically $\sim$\,180 km from the detector. Suppose the LMA-MSW effect is indeed the mechanism responsible for solar neutrino flavor change. 
Then, assuming CPT invariance, the neutrino and antineutrino properties are such that KamLAND should see substantial disappearance of reactor $\overline{\nu_e}$ flux. On the other hand, if some other version of the MSW effect or vacuum oscillation is behind solar neutrino behavior, then KamLAND should see an undiminished flux. 
What KamLAND actually does see is a flux only 0.611 $\pm$ 0.085 (stat) $\pm$ 0.041 (syst) of the value it would have in the absence of disappearance \cite{r11}. The KamLAND rate and spectral observations and the solar neutrino data have now been compared by many authors to the neutrino parameters corresponding to the various standard explanations of solar neutrino behavior \cite{r11,r12}. 
It is found that parameters corresponding to the LMA-MSW effect fit both the solar and KamLAND data very well. In contrast, the parameters required by any other standard explanation of solar neutrino behavior are ruled out with high confidence. The LMA-MSW effect has been uniquely identified as the mechanism underlying solar neutrino flavor change.

For CP violation in neutrino oscillation to be visible in terrestrial experiments, the neutrino mass splittings and mixing angles must be sufficiently large. If there are only three neutrinos, all of these quantities, including those pertaining to the solar neutrinos, must be of sufficient size. 
Among the once viable candidate solar-flavor-change mechanisms, only the LMA-MSW effect involves a mass splitting and a mixing angle that are both adequate. Thus, the demonstration by KamLAND that LMA-MSW is indeed the candidate that has been chosen by Nature is very good news indeed.

The evidence that solar neutrinos change flavor joins earlier convincing evidence that the atmospheric neutrinos do so as well. The latter evidence  includes the observation that, for multi-GeV atmospheric muon neutrinos observed in the SK detector \cite{r13},
\beq
	\frac{\mbox {Flux Up} {\scriptstyle (-1.0 < \cos \theta_Z < -0.2)}}
	{\mbox {Flux Down} {\scriptstyle (+0.2 < \cos \theta_Z < +1.0)}} = 0.54 \pm 0.045 \,\,  .
\label{eq2.5}
\eeq
Here, $\theta_Z$ is the zenith angle of the incoming neutrinos, with $\cos \theta_Z = +1$ corresponding to vertically downgoing neutrinos and $\cos \theta_Z = -1$ to vertically upcoming ones. Owing to the observed isotropy of the cosmic rays that produce the multi-GeV atmospheric neutrinos, the Up/Down ratio in \Eq{2.5} must be unity, unless some of the produced muon neutrinos disappear (or else additional ones appear) between their production in the atmosphere and their detection in the detector \cite{r14}. 
Thus, the observed Up/Down ratio implies that some mechanism must be changing the $\nu_\mu$ flux as the neutrinos travel to the detector. Voluminous data from several detectors are beautiffuly described by the hypothesis that this mechanism is the oscillation $\nu_\mu \to \nu_?$ of muon neutrinos into neutrinos $\nu_?$ of another flavor \cite{r15}. 
Since the upward-going $\nu_\mu$ are created in the atmosphere on the far side of the earth from the detector, they have much more time to oscillate away into $\nu_?$ while enroute than do the downward-going $\nu_\mu$, which are created in the atmosphere right above the detector. This explains why FluxUp $<$ Flux Down. 
The data (including reactor data) tell us that $\nu_?$ is at least mostly $\nu_\tau$, that the $\nu_\mu - \nu_?$ mixing is very large and perhaps maximal, and that the oscillation reflects a neutrino (mass)$^2$ difference $\Delta m^2_{\mathrm{atm}} \sim 2.5 \times 10^{-3}\,$ eV$^2$.

The detailed comparison between the atmospheric neutrino data and the oscillation hypothesis depends, of course, on a theoretical knowledge of what the neutrino fluxes would be if there were no oscillation. At lower energies, these fluxes are azimuthally distorted by geomagnetic effects. 
The theoretical predictions for this distortion have been confirmed by flux measurements that are largely independent of oscillation effects \cite{r16}. This further strengthens one's confidence that the observed deviations between the atmospheric neutrino fluxes and the no-oscillation predictions are indeed due to oscillations.

Needless to say, it is highly desirable to verify that atmospheric neutrinos are undergoing the oscillation $\nu_\mu \to \nu_\tau$ with a mass splitting $\Delta m^2_{\mathrm{atm}} \sim 2.5 \times 10^{-3}$ eV$^2$ and $\sim$ maximal mixing by showing that \emph{accelerator-produced} $\nu_\mu$ undergo the same oscillation, with the same parameters. To this end, several experiments that allow accelerator neutrinos to traverse a Long Base Line $L$ are in progress or under construction. 
The K2K experiment, with $L$ = 250\,km, has already reported results. At its typical neutrino energy $E$ of $\sim$\,1.3 GeV, this experiment has $\sin^2 \,[1.27 \Delta m^2_{\mathrm{atm}}$(eV$^2) L$(km)/$E$(GeV)], the characteristic factor governing the probability of the expected oscillation, equal to 1/3, so that significant oscillation should be seen. 
In the results reported at Neutrino 2002 and in a December, 2002 paper \cite{r18}, K2K observes 56 $\nu_\mu$ events in the detector at the far end of the 250 km baseline. However, based on $\nu_\mu$ flux measurements by the K2K near detectors, there should have been 80 events in the far detector if there were no oscillation. In addition, K2K was able to get some information on the energy dependence of the apparent oscillation. 
Using all of its information, it found the probability of no oscillation to be less than 1\%. All of its data are successfully described assuming oscillation, and, interestingly enough, they are fit best by a mass splitting $\Delta m^2_{\mathrm{K2K}} \sim 2.8 \times 10^{-3}$ eV$^2$ and a mixing parameter $(sin^2\,2\theta)_{\mathrm{K2K}} = 1.0$ that agree almost perfectly with the parameters $\Delta m^2_{\mathrm{atm}} \sim 2.5 \times 10^{-3}$ eV$^2$ and $(sin^2\,2\theta)_{\mathrm{atm}} = 1.0$ that give the best fit to the SK atmospheric neutrino data.

\vspace{1pc}
{\bf The Future}

One would like to further confirm the $\nu_\mu \to \nu_\tau$ oscillation of atmospheric neutrinos by strengthening the evidence for the same oscillation by accelerator neutrinos. The K2K experiment, after an unfortunate interruption, is already contributing to this goal. The MINOS \cite{r19} and CNGS experiments, under construction, could greatly enhance the evidence.

In vacuum, or in matter of negligible influence, the probability for a neutrino flavor change has a characteristic undulatory dependence on $L/E$ of the form $\sin^2 \,[1.27 \Delta m^2$(eV$^2) L$(km)/$E$(GeV)]. Indeed, that is why this flavor change is called ``oscillation''. While the evidence for flavor change is already quite convincing, the characteristic undulation has not yet been observed. 
Since this undulation is such a central feature of vacuum oscillation, it is obviously very important to actually see it. With further running, the KamLAND experiment can perhaps go some way toward this goal with respect to the reactor antineutrinos. 
Hopefully, the future Long Base Line (LBL) neutrino experiments will be able to observe the undulation of accelerator neutrino oscillation. Perhaps future underground detectors will even be able to see the undulation of atmospheric neutrino oscillation \cite{r21}.

\section{HOW MANY NEUTRINO SPECIES ARE THERE?}

The solar, atmospheric, and LSND oscillations call, respectively, for (mass)$^2$ splittings $\Delta m^2$ of $\sim$\,Few $\times 10^{-5}\,$eV$^2$, $\sim$\,Few $\times 10^{-3}\,$eV$^2$, and $\sim$\,1 eV$^2$. Now, if there are only three mass eigenstates $\nu_i$, each with a mass $m_i$, then there are only three possible (mass)$^2$ splittings $\Delta m^2_{ij} \equiv m^2_i - m^2_j$, and these three splittings obviously satisfy the relation $\Delta m^2_{32} + \Delta m^2_{21} + \Delta m^2_{13} = 0$. 
Clearly, the three splittings required by the reported solar, atmospheric, and LSND oscillations do not obey this constraint. Thus, given the strong evidence for the solar and atmospheric oscillations, if the so-far unconfirmed LSND oscillation is confirmed as well, then Nature must contain at least four neutrino masses. 
These are either the masses of four different mass eigenstates $\nu_i$, or else CPT invariance is broken in such a way that mass$\,(\overline{\nu_i}) \neq$ mass$\,(\nu_i)$, so that there can be six distinct masses from three $\nu_i$ plus three $\overline{\nu_i}$ \cite{r23}. (See Sec. 8) 
If CPT is not broken and there are, say, four $\nu_i$, then one linear combination of them, $\nu_s$, has no charged lepton partner (there being only three charged leptons: e, $\mu$, and $\tau$), and consequently cannot couple to the $W$. From the observed fact that the decays $Z \to \nu\bar{\nu}$ yield only three distinct neutrino species, $\nu_s$ evidently does not couple to the $Z$ either. Thus, $\nu_s$ is called a ``sterile'' neutrino.

We see that confirmation of LSND would imply a very new phenomenon---either a sterile neutrino, or CPT violation. Either would be extremely interesting.

\vspace{1pc}
{\bf The Future}

The MiniBooNE experiment at Fermilab is aimed at confirming or refuting LSND. MiniBooNE is already taking data. Its result will be very important, to say the least.

\section{WHAT IS THE NEUTRINO MASS SPECTRUM? 
              WHAT IS THE MIXING MATRIX $U$?}
              
\begin{figure}[htb]
\includegraphics[width=7cm]{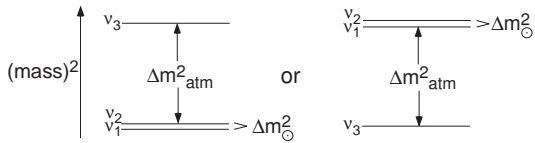}
\caption{The neutrino (mass)$^2$ spectrum assuming only three neutrinos.}
\label{f1}
\end{figure}

If only the atmospheric and solar neutrino oscillations prove to be genuine and LSND is not confirmed, then Nature may contain only three mass eigenstates. The (mass)$^2$ spectrum then has the character shown in Fig.~\ref{f1}. 
There are two eigenstates (the ``solar pair'') separated from each other by the small splitting $\Delta m^2_\odot (\sim 7 \times 10^{-5}\,$eV$^2)$ that drives solar neutrino flavor change. (The symbol $\odot$ is the astronomers' symbol for the sun.) There is also a third eigenstate separated from the solar pair by the larger splitting $\Delta m^2_{\mathrm{atm}} (\sim 2.5 \times 10^{-3}\,$ eV$^2$) that drives atmospheric neutrino flavor change. 
This isolated eigenstate, $\nu_3$, may be either heavier or lighter than the solar pair, as shown in Fig.~\ref{f1}. We can find out which of these is the case by studying flavor change in an accelerator-generated neutrino beam that traverses a L(ong) B(ase) L(ine) through earth matter. In such an experiment, one is beating the sign one wishes to determine, that of mass$^2\,(\nu_3)$ - mass$^2\,$(solar pair), against a sign one knows, that of the extra energy electron neutrinos acquire through their coherent forward scattering from ambient electrons as they pass through matter. 
The latter sign is positive. The principle behind this sign detemination is similar to that behind the determination of the sign of mass ($K_L$) - mass ($K_S$). The latter is found by passing kaons through matter known as a regenerator, and beating the sign one wishes to determine against the known sign of the regeneration amplitude.

If there are only three neutrinos, then the mixing matrix $U$ is 3 $\times$ 3, and from what we have already learned about neutrino flavor change, this matrix is given approximately by \cite{r24}
\begin{eqnarray}  
\lefteqn{U = }\hspace{-0.5cm}  \nonumber  \\
  &  &  \hbox{\hskip1.2cm}\nu_1  \hbox{\hskip2.0cm} 
       \nu_2  \hbox{\hskip1.5cm} \nu_3     \nonumber  \\
  &  &  \hbox{\hskip-0.5cm}\begin{array}{c}
 	\nu_e  \\  \nu_\mu \\ \nu_\tau 
 		\end{array}
\hspace{-0.25cm}\left[  \hspace{-0.2cm}
\begin{array}{ccc}
	\phantom{-}c\,e^{i\alpha_1/2} 
         & \phantom{-}s\,e^{i\alpha_2/2} & s_{13}\,e^{-i\delta} \\
    -s\,e^{i\alpha_1/2}/\sqrt{2} 
         & \phantom{-}c\,e^{i\alpha_2/2}/\sqrt{2} & 1/\sqrt{2} \\
    \phantom{-}s\,e^{i\alpha_1/2}/\sqrt{2} 
         & -c\,e^{i\alpha_2/2}/\sqrt{2} & 1/\sqrt{2} \\ 
\end{array}
\hspace{-0.18cm}  \right]
\label{eq4.1} 
\end{eqnarray}
Here, the symbols outside the matrix label its rows and columns, and, as in Fig. \ref{f1}, $\nu_3$ is the isolated mass eigenstate, regardless of whether it is heavier or lighter than the other two. In writing the matrix of \Eq{4.1}, we have assumed that the atmospheric neutrino mixing is maximal. We have introduced $c \equiv \cos\theta_\odot$ and $s \equiv \sin\theta_\odot$, where $\theta_\odot$ is the large solar mixing angle inferred from the LMA-MSW explanation of solar neutrino behavior. At 90\% confidence level \cite{r25},
\begin{eqnarray*}
0.25 & \lsim & \sin^2 \,\theta_\odot \lsim 0.40 ~~. 
\end{eqnarray*}
We have also introduced $s_{13} \equiv \sin\theta_{13}$, where $\theta_{13}$ is a mixing angle known to be small from bounds on Short Base Line reactor $\overline{\nu_e}$ oscillation. At 90\% confidence level \cite{r26},
\beq
\sin^2\,\theta_{13} \lsim 0.03~~.
\label{eq4.2}
\eeq
Finally, we have introduced $\delta,\; \alpha_1$, and $\alpha_2$, which are CP-violating phases whose values are unknown. 

The character of the leptonic mixing matrix $U$ of \Eq{4.1} is a big surprise. The quark analogue of $U$, $V$, has the structure
\beq
V = \left[  \begin{array}{ccc}
	1 & s & s \\
	s & 1 & s \\
	s & s & 1 
		\end{array} \right]  ~~,
\label{eq4.3}
\eeq
where ``$s$'' denotes an entry that is small compared to unity. It was natural to expect that $U$ would look similar. But in reality it looks very different, having the structure
\beq
U = \left[  \begin{array}{ccc}
	B & B & s \\
	B & B & B \\
	B & B & B 
		\end{array} \right]  ~~.
\label{eq4.4}
\eeq
Here ``$B$'' stands for an entry that is big; that is, an appreciable fraction of unity.

The striking contrast between $V$, with its small mixing angles, and $U$, with its large ones, probably contains a clue to the physics that underlies mixing. However, we do not yet know how to interpret this clue.

\begin{figure}[htb]
\includegraphics[width=7cm]{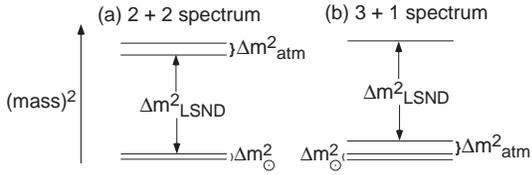}
\caption{Possible four-neutrino (mass)$^2$ spectra.}
\label{f2}
\end{figure}

If not only the atmospheric and solar neutrino oscillations but also the LSND one prove to be genuine, then Nature must contain at least four neutrino masses, as we have explained. Assuming CPT invariance, these masses are the masses of four neutrino mass eigenstates $\nu_i$. If there are four $\nu_i$ , the spectrum has the character shown in Fig.~\ref{f2}(a) or Fig.~\ref{f2}(b).

In Fig.~\ref{f2}(a), we have a ``2 + 2'' spectrum. This is comprised of a ``solar pair'' of eigenstates that are the major contributors to the behavior of solar neutrinos and are separated from each other by the solar splitting $\Delta m^2_{\odot}$, plus an ``atmospheric pair'' that are the main contributors to the atmospheric $\nu_\mu \to \nu_\tau$ oscillation and are separated by the atmospheric splitting $\Delta m^2_{\mathrm{atm}}$. The solar pair is separated from the atmospheric pair by the large splitting $\Delta m^2_{\mathrm{LSND}} \; (\sim 1\,$eV$^2)$ required by the LSND oscillation. The solar pair may lie below the atmospheric pair as shown, or above it.

From bounds on $\overline{\nu_e}$ \cite{r27} and $\nu_\mu$ \cite{r28} oscillation over suitably short base lines, we know that the $\nu_e\; (\nu_\mu)$ fraction of the atmospheric (solar) pair is not much more than 3\%. If these small fractions are neglected, then the 2 + 2 spectrum requires that sterile neutrino production in the atmospheric and/or solar oscillations be substantial. In particular, this production must satisfy the sum rule \cite{r29}
\beq
f_s^{\mathrm{atm}} + f_s^\odot = 1 ~~.
\label{eq4.5}
\eeq
Here,
\beq 
f_s^{\mathrm{atm}} \equiv \left.  \frac{P(\nu_\mu \to \nu_s)} 
	{P(\nu_\mu \to \nu_\tau) + P(\nu_\mu \to \nu_s)}
	\right|_{\mathrm{atmos.} \atop \mathrm{neutrinos}} ,
\label{eq4.6}
\eeq
and
\beq
 f_s^\odot \equiv \left.  \frac{P(\nu_e \to \nu_s)} 
	{P(\nu_e \to \nu_\tau) + P(\nu_e \to \nu_s)}
	\right|_{\mathrm{solar} \atop \mathrm{neutrinos}} ,
\label{eq4.7}
\eeq
with $\nu_s$ a sterile neutrino and $P$ an oscillation probability. (In the approximation we are making, atmospheric $\nu_\mu$ do not oscillate to $\nu_e$, and solar $\nu_e$ do not oscillate to $\nu_\mu$.) Experimentally, $f_s^{\mathrm{atm}} < 0.19$ \cite{r15}, and $f_s^\odot < 0.36$ \cite{r30}, both at 90\% confidence level. Thus, it appears that the sum rule of \Eq{4.5} is probably not obeyed. 
However, it has been discovered that when the small $\nu_e \,(\nu_\mu)$ fraction of the atmospheric (solar) pair is not neglected, and certain matter effects are included, large deviations from this sum rule are allowed, even if the underlying neutrino spectrum is of the 2 + 2 variety \cite{r31}. Thus, the existing data and analyses do not exclude the possibility of a 2 + 2 spectrum.

In Fig.~\ref{f2}(b), we have a ``3 + 1'' spectrum. This consists of a trio, made up of a solar pair separated by $\Delta m^2_{\odot}$ and a third neutrino separated from the solar pair by $\Delta m^2_{\mathrm{atm}}$, plus a fourth neutrino separated from the trio by $\Delta m^2_{\mathrm{LSND}}$. In the trio, the solar pair may be at the top or bottom, and the entire trio may be above or below the fourth neutrino.

In a 3 + 1 spectrum, essentially all the sterile flavor content can be placed in the fourth, isolated, neutrino. The solar and atmospheric oscillations are then driven by the neutrinos in the trio. Since these are almost completely active, the solar and atmospheric oscillations will produce almost no sterile flux, so even very tight upper bounds on $ f_s^\odot$ and $f_s^{\mathrm{atm}}$ would not be a problem.
 However, in a 3 + 1 spectrum, the probability of the LSND $\overline{\nu_\mu} \to \overline{\nu_e}$ oscillation is proportional to $|U_{\mu4}\, U_{e4}|^2$, where $U$ is now a 4$\times$4 mixing matrix, and mass eigenstate $\nu_4$ is the isolated neutrino in the spectrum---the ``1'' of 3 + 1. 
The value of $|U_{\mu4}\, U_{e4}|^2$ favored by LSND is somewhat large relative to the upper bounds on $U_{\mu4}$ and $U_{e4}$ from null oscillation searches. Thus, the 3 + 1 spectra are not a great fit to all the data \cite{r32}, but they are not excluded.

The conclusion from this consideration of spectra with more than three mass eigenstates is that the LSND oscillation signal is alive, and whether it is genuine needs to be settled experimentally. Thus, the MiniBooNE experiment, indended precisely to settle this issue, is indeed crucially important.

\vspace{1pc}
{\bf The Future}

One would like to determine the (mass)$^2$ splittings $\Delta m^2_{\odot},\; \Delta m^2_{\mathrm{atm}}$, and $\Delta m^2_{\mathrm{LSND}}$, and the corresponding mixing angles $\theta_\odot,\; \theta_{\mathrm{atm}}$, and $\theta_{\mathrm{LSND}}$, more precisely than they are known at present. 
For example, one would like to know whether the atmospheric mixing angle $\theta_{\mathrm{atm}}$ is truly maximal ($\sin^2\,2\theta_{\mathrm{atm}} = 1$), representing maximal mixing between $\nu_\mu$ and $\nu_\tau$, or deviates somewhat from maximality. 
The observed maximal or near-maximal mixing suggests the presence of a symmetry, and a deviation from maximality would then reflect the breaking of this symmetry. We may describe this situation by writing the $\nu_\mu - \nu_\tau$ mass matrix (neglecting mixing with $\nu_e$) in the form
\begin{eqnarray}
 & & \hspace{0.9cm}\nu_\mu \hspace{0.3cm}\nu_\tau  \hspace{1.7cm}\nu_\mu \hspace{0.3cm}\nu_\tau \nonumber\\
\mathcal{M} & = & \hspace{-0.1cm} \begin{array}{c}  \nu_\mu \\  \nu_\tau   \end{array}  \hspace{-0.15cm}
	\left[  \begin{array}{cc} m  &  x  \\  x  &  m  \end{array} \right]  + 
	\begin{array}{c}  \nu_\mu \\ \nu_\tau  \end{array}  \hspace{-0.15cm}
	\left[  \begin{array}{cc} \delta  & 0 \\  0  &  -\delta  
	\end{array}  \right] ~, 
\label{eq4.8}
\end{eqnarray}
where the symbols outside the matrices label the rows and columns, and $\delta \ll m,\;x$. The first matrix in \Eq{4.8} is symmetric, and when $\delta = 0$ leads to maximal mixing. The second matrix breaks the symmetry, and leads to a deviation from maximal mixing given by 
\beq
1 - \sin^2\,2\theta_{\mathrm{atm} }\cong \left( \frac{\delta}{x} \right)^2 ~~.
\label{eq4.9}
\eeq
Thus, the deviation $1 - \sin^2\,2\theta_{\mathrm{atm}}$ measures the scale of symmetry breaking.

It would be very nice if theory could provide a guide as to how precisely the various (mass)$^2$ splittings and mixing angles need to be known.

\begin{figure}[htb]
\includegraphics[width=6cm]{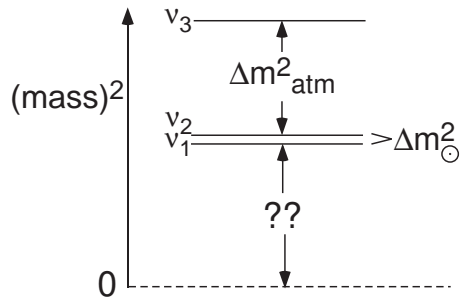}
\caption{A schematic (mass)$^2$ spectrum emphasizing the importance of its height above zero.}
\label{f3}
\end{figure}

One very important question is how high the entire neutrino (mass)$^2$ spectrum lies above zero. If, for example, the spectrum has the form shown in Fig.~\ref{f3}, we would like to know the size of the gap labelled ``??'', so that we will know the absolute scale of neutrino mass.
To develop a theoretical understanding of neutrino masses, we need to know how big these masses are, and not just relative to one another.

Determining the absolute scale of neutrino mass is a daunting challenge. One approach is to study the $\beta$ spectrum in tritium decay \cite{r33}. The forthcoming KATRIN tritium experiment would be able to detect the mass $m_i$ of a mass eigenstate $\nu_i$ if $m_i \gsim $0.5 eV {\em and} $\nu_i$ couples appreciably to an electron. Interestingly, if the LSND oscillation is genuine, then there must be a $\nu_i$ whose mass exceeds $\sqrt{\Delta m^2_{\mathrm{LSND}}}$. Given the favored range for $\sqrt{\Delta m^2_{\mathrm{LSND}}}$, the mass of this $\nu_i$ exceeds $\sim\,$0.4 eV. If this $\nu_i$ couples appreciably to an electron, its mass could be within range of KATRIN \cite{r34}.

Another approach is to try to determine the effect of neutrino mass on cosmology \cite{r35}. Ideally, we would like cosmological probes to be sensitive to a neutrino mass as small as 0.05 eV, since the convincingly established atmospheric oscillation tells us that there is a mass eigenstate whose mass is at least  $\sqrt{\Delta m^2_{\mathrm{atm}}} \cong 0.05\,$eV. 
In an intriguing post-Neutrino 2002 paper, it is suggested that weak gravitational lensing experiments could perhaps determine a mass this small \cite{r36}. However, the neutrino mass reach of these experiments depends on a knowledge of other things, such as the dark energy density.

Combining the just-announced cosmic microwave background results from the Wilkinson Microwave Anisotropy Probe (WMAP) with earlier results, it is found that at 95\% CL \cite{r36aMAP},
\beq
\sum_i m_i < 0.71\, \mathrm{eV} ~~,
\label{eq4.10}
\eeq
where, as usual, $m_i$ is the mass of $\nu_i$. This bound, tighter than the present one from tritium $\beta$ decay, is a very interesting bound indeed.

\section{DOES $\bar{\nu} = \nu$?}

As emphasized by Yanagida at Neutrino 2002, it is a generic prediction of the see-saw mechanism that each neutrino mass eigenstate $\nu_i$ is identical to its antiparticle $\bar{\nu}_i$. As emphasized by Valle, the observation of neutrinoless double beta decay ($0\nu\beta\beta$) at any nonzero level would confirm that indeed $\bar{\nu}_i = \nu_i$. 
(To be sure, this statement assumes CPT invariance \cite{r37}.) Neutrinoless double beta decay is the decay Nucl $\to$ Nucl$^\prime$ + 2e$^-$ of one nucleus into another plus two electrons. The amplitude for this process is proportional to the quantity
\beq
m_{\beta\beta} \equiv | \sum_i m_i\, U^2_{ei} | ~~.
\label{eq5.1}
\eeq 
Clearly, $m_{\beta\beta} $---the effective neutrino mass for $0\nu\beta\beta$---is a measure of the neutrino mass scale. Thus, the observation of $0\nu\beta\beta$, confirming that neutrinos are identical to their antiparticles, could also tell us about the mass scale. The desirable sensitivity of  $0\nu\beta\beta$ experiments is to $m_{\beta\beta} \lsim 0.05\,$eV, since as already noted, the atmospheric oscillation tells us that one mass $m_i$ in \Eq{5.1} is at least 0.05 eV.

Suppose that the neutrino mass eigenstates are indeed identical to their antiparticles, and that the neutrino spectrum has the form depicted on the right-hand side of Fig.~\ref{f1}. Then $0\nu\beta\beta$ occurs, and the contribution of the solar pair, $\nu_{1,2}$, dominates $m_{\beta\beta}$, both because $|U_{e3}|^2 = \sin^2 \theta_{13} \lsim 0.03$ [cf. \Eq{4.2}], and because $\nu_3$ is lighter than the pair. 
Taking into account that a practical $0\nu\beta\beta$ experiment cannot see the tiny splitting between the members of the solar pair,
\beq
m_{\beta\beta} \cong m_0 \sqrt{1 - \sin^2 2\theta_\odot \sin^2  \left( \frac{\alpha_2 - \alpha_1}{2}  \right) } ~~.
\label{eq5.2}
\eeq
Here, $m_0$ is the average mass of the members of the solar pair, $\theta_\odot$ is the solar mixing angle, and $\alpha_{1,2}$ are the phases that appear in the mixing matrix of \Eq{4.1}. We note from \Eq{5.2} that, whatever the values of  $\alpha_{1,2}$,
\beq
m_{\beta\beta} \geq m_0 \cos2\theta_\odot ~~.
\label{eq5.3}
\eeq
Using $m_0 \geq \sqrt{\Delta m^2_{\mathrm{atm}}} $, and taking for $\Delta m^2_{\mathrm{atm}}$ and $\cos2\theta_\odot$ the best fit values of $2.5 \times 10^{-3}\,$eV$^2$ \cite{r15} and 0.38 \cite{r25}, respectively, this relation implies that $m_{\beta\beta} \geq 0.019\,$eV. 
Even with generous allowances for the uncertainties, one finds \cite{r38} that $m_{\beta\beta} \gsim 0.0085$ eV. Clearly, the planned $0\nu\beta\beta$ experiments with $m_{\beta\beta}$ sensitivities in the 0.01 eV to 0.1 eV range \cite{r39} could prove to be very interesting.

An  intriguing paper reporting evidence for $0\nu\beta\beta$ \cite{r40} has already appeared. The reported evidence has led to considerable discussion, and remains controversial. We eagerly await the results of future $0\nu\beta\beta$ experiments.

\section{DOES NEUTRINO BEHAVIOR VIOLATE CP?}

The observations of CP violation in the behavior of neutrinos would establish that CP violation is not a peculiarity of quarks, but occurs among leptons as well. If baryogenesis in the early universe came about through leptogenesis followed by the conversion of a lepton asymmetry into a baryon asymmetry, then leptonic CP violation exists. Leptogenesis is impossible without it. 
To be sure, the leptonic CP violation that made leptogenesis possible, and the one that we might see in neutrino oscillation are independent phenomena whose relation to each other is model-dependent. Nevertheless, if leptogenesis occurred, then it is likely that CP violation in neutrino oscillation occurs as well \cite{r41}.

If there are only three neutrinos, then the leptonic mixing matrix can contain the three CP-violating phases $\delta,\;\alpha_1$, and $\alpha_2$ that appear in \Eq{4.1}. It is easily shown that $\delta$, and only $\delta$, can lead to CP violation in neutrino oscillation. 
This CP violation would manifest itself as a CP-violating difference $\Delta_{CP}(\alpha\beta)$ between the probability $P(\nu_\alpha \to \nu_\beta)$ for oscillation of a neutrino of flavor $\alpha$ into one of flavor $\beta$ and the probability $P(\overline{\nu_\alpha} \to \overline{\nu_\beta})$ for the corresponding antineutrino oscillation \cite{r42}:
\beq
\Delta_{CP}(\alpha\beta) \equiv  P(\nu_\alpha \to \nu_\beta) - P(\overline{\nu_\alpha} \to \overline{\nu_\beta}) ~~.
\label{eq6.1}
\eeq
If there are only three neutrino flavors, then there are only three independent CP-violating differences $\Delta_{CP}(\alpha\beta)$ that one can measure: $\Delta_{CP}(e\mu),\; \Delta_{CP}(\mu\tau)$, and $\Delta_{CP}(\tau e)$. Interestingly enough, it follows from the general expressions for $P(\optbar{\nu_\alpha} \to \optbar{\nu_\beta})$ when matter effects are negligible that
\begin{eqnarray}
\Delta_{CP}(e\mu)&  = & \Delta_{CP}(\mu\tau)\; = \;\Delta_{CP}(\tau e) \nonumber \\
	& = & 16 J\, k_{12}k_{23}k_{31} ~~.
\label{eq6.2}
\end{eqnarray}
Here,
\begin{eqnarray}
J & \equiv & \Im (U^*_{e1}U_{e3}U_{\mu1}U^*_{\mu3}) \nonumber \\
	& \cong & \frac{1}{4} \sin 2\theta_\odot \sin\theta_{13}\sin\delta ~~,
\label{eq6.3}
\end{eqnarray}
and
\beq
k_{ij} \equiv \sin [1.27 \Delta m^2_{ij} (\mbox{eV}^2) \frac{L(\mbox{km})}{E(\mbox{GeV})}  ]~.
\label{eq6.4}
\eeq
As we see, the predicted CP violation is beautifully simple. All three CP-violating differences $\Delta_{CP}(\alpha\beta)$ are equal, and their common value reflects the underlying neutrino parameters---the mixing angles, phase, and (mass)$^2$ splittings---in a way that is completely free of the hadronic uncertainties that sometimes bedevil efforts to interpret CP-violating effects in hadronic decay. 
To be sure, the predicted $\Delta_{CP}(\alpha\beta)$ in neutrino oscillation is small---perhaps of order (1-2)\%---due to the small size of $\theta_{13}$ and of the smallest of the $\Delta m^2_{ij}$, which is the solar splitting $\Delta m^2_\odot$.

If there are more than three neutrinos, then the possibilities for CP violation in oscillation become very rich.

As \Eq{4.1} makes clear, all effects of the phase $\delta$ are proportional to $\sin\theta_{13}$. Equation (\ref{eq6.3}) confirms that this is true for CP violation in oscillation. At present, we know only that $\sin\theta_{13} \lsim 0.2$ [cf. \Eq{4.2}]. It is obviously very important to demonstrate experimentally that $\theta_{13}$ does not vanish, and to find out how large it is. A number of Long Base Line future experimental programs that are under discussion have the measurement of $\theta_{13}$ as a major goal.

In practice, neutrino and antineutrino oscillation probabilities depend on several factors at once: genuine CP violation from the phase $\delta$, neutrino-antineutrino asymmetries induced by the passage of the neutrinos through earth matter that is not CP symmetric, and CP-conserving neutrino parameters. It will be necessary to make a number of complementary measurements and analyze them jointly to disentangle the various neutrino properties \cite{r43}.

The CP-violating phases $\alpha_1$ and $\alpha_2$, known as Majorana phases, have physical consequences only if the neutrinos are Majorana particles. As previously mentioned, they do not affect neutrino oscillation (regardless of the character of the neutrinos). However, if neutrinos are Majorana particles, then as we see from \Eq{5.1} and the $U$ matrix, \Eq{4.1}, the phases $\alpha_{1,2}$ do affect the rate for $0\nu\beta\beta$ by influencing $m_{\beta\beta}$. 
Actually detecting the presence of $\alpha_{1,2}$ through a measurement of the rate for $0\nu\beta\beta$ would require good fortune. In an optimistic scenario, the neutrino spectrum has the form shown on the right-hand side of Fig.~\ref{f2}, and the mass eigenstates in the solar pair have an average mass $m_0 \sim 0.5\,$eV, large enough to be measured in the forthcoming tritium $\beta$ decay experiment KATRIN. 
With $m_0$ and $\theta_\odot$ known, \Eq{5.2} for $m_{\beta\beta}$ determines $\alpha_2 - \alpha_1$ once $m_{\beta\beta}$ has been determined. However, the uncertainties in $m_0$ and $\theta_\odot$ combine with a potentially large uncertainty in $m_{\beta\beta}$, from both experimental and theoretical sources, to make the determination of $\alpha_2 - \alpha_1$, or even the demonstration that this relative CP-violating phase is nonvanishing, problematical. 
It has been argued that these goals are unreachable \cite{r44}. However, somewhat less pessimistic assessments have been made \cite{r45}.

We note that the scenario we have described, with $m_0 \sim 0.5\,$eV, is in some conflict with the new WMAP bound of \Eq{4.10}. However, perhaps future tritium experiments would be able to measure somewhat smaller masses than this.

\section{WAS BARYOGENESIS MADE POSSIBLE BY LEPTONIC CP VIOLATION?}

The present-day universe contains an excess of baryons over antibaryons. Symmetry suggests that there was no such excess at the time of the big bang. But the subsequent development of the excess---baryogenesis---could not have occurred without CP violation, and it is known that CP violation arising from the CP violating phase in the {\em quark} mixing matrix would not have been nearly sufficient. 
As a result, there is growing interest in the possibility that, through a two-step process, the baryon excess grew out of {\em leptonic} CP violation. The first step in this process was leptogenesis---the production of an excess of antileptons over leptons. The second step was the conversion of this antilepton excess into a baryon excess through nonperturbative Standard Model B-L conserving processes.

In the see-saw mechanism, a Dirac neutrino gets split by Majorana mass terms into two Majorana particles: a very light one identified as one of the light neutrinos, and a very heavy neutral lepton, $N$. It is thought that leptogenesis may have been the result of a CP-violating difference between  the rates for decay of the Majorana particle $N$ into leptons and antileptons \cite{r46A}:
\beq
\Gamma (N \to \ell^+ + H^-) > \Gamma (N \to \ell^- + H^+) ~~.
\label{eq7.1}
\eeq
Here $\ell^-$ is a charged lepton, and $H^+$ is the charged Higgs boson. (When the electroweak phase transition occurs, $H^+$ will become the longitudinal state of the $W^+$, but at earlier times it is an ordinary particle.) Clearly, the CP-violating difference of rates in \Eq{7.1} leads to an antilepton excess.

As noted in Sec. 6, the relation between the leptonic CP-violation that leads to the inequality of \Eq{7.1} and the one that we might observe in neutrino oscillation is model-dependent. Nevertheless, demonstrating that leptonic CP violation does exist by observing it in neutrino oscillation would increase our confidence that leptogenesis may indeed have led to the baryon asymmetry of the universe.

\section{DO NEUTRINOS VIOLATE CPT?}

It has been speculated that space may have extra dimensions, beyond the immediately visible three, and that only particles devoid of nontrivial Standard Model quantum numbers can travel in the extra dimensions. The candidate particles are the gravitons and the right-handed neutrinos. While traveling in the extra dimensions, the right-handed neutrinos might encounter CPT-violating effects from string-induced structure. 
Conceivably, this could lead to a large CPT-violating difference between mass ($\nu_i$) and mass ($\overline{\nu_i}$). In turn, this CPT-violating decoupling of the neutrino and antineutrino mass spectra would lead to CPT-violating differences between neutrino oscillations and antineutrino oscillations. The observation that the oscillation of solar {\em neutrinos} and reactor {\em antineutrinos} are both successfully described by a common set of parameters would appear to be evidence against large-scale CPT violation by neutrinos. 
However, it has been pointed out that present data still allow for considerable CPT violation, and that they can all be described in terms of a CPT-violating pair of spectra---one for neutrinos and one for antineutrinos---with only three states in each spectrum \cite{r47B}.

\section{WHAT CAN NEUTRINOS TELL US ABOUT ASTROPHYSICS AND COSMOLOGY, OR VICE VERSA?}

Neutrinos offer a potentially very interesting window on the universe. Through the elastic scattering of supernova neutrinos from protons, the total energy and temperature of supernova $\nu_\mu$ and $\nu_\tau$ can be found \cite{r48C}. As already noted in Sec. 4, astrophysical and cosmological observations can provide important information about the properties of the neutrinos.

\section{WHAT IS THE ORIGIN OF NEUTRINO FLAVOR PHYSICS?}

The ultimate question is wfhat lies behind the rich phenomena we are discovering in  neutrino physics. The see-saw mechanism, and other simple arguments, lead us to expect that Majorana masses---masses that mix neutrinos and antineutrinos, leading to mass eigenstates that are their own antiparticles---are involved. The see-saw mechanism leads to the see-saw relation
\beq
m_\nu = \frac{m^2_{\mathrm{quark}}} {m_{\mathrm{large}}} ~~,
\label{eq10.1}
\eeq
where $m_\nu$ is some neutrino mass, $m_{\mathrm{quark}}$ some quark mass, and $m_{\mathrm{large}}$ a very large mass. Given the lightness of neutrinos, this relation implies that $m_{\mathrm{large}}$ is large indeed, suggesting that physics at a very high mass scale is responsible for the neutrino masses. 
However, another, equally interesting, possibility is that neutrino masses are Dirac masses, which are of the form $\overline{\nu_L} \nu_R$, and that these masses are small because the right-handed neutrino $\nu_R$ is ``lost'' in an extra dimension of space, so that it has small overlap with the left-handed neutrino $\nu_L$, which is confined to our three-dimensional world \cite{r49E}. 

Surprisingly, two of the neutrino mixing angles have turned out to be large. Are symmetries behind this? The roles of symmetries and naturalness in neutrino mass and mixing were reviewed at Neutrino 2002 by King \cite{r50D}. Neutrino mass models were also explored by Valle \cite{r49E}.

\section{WHAT ARE THE CONNECTIONS BETWEEN NEUTRINO AND QUARK FLAVOR PHYSICS?}

In G(rand) U(nified) T(heories), where quarks and leptons are in the same family, one expects quark and leptonic flavor physics to be related to each other. It has been pointed out that in a supersymmetric SU(5) GUT, the $\sim$ maximal $\nu_\mu -\nu_\tau$ mixing inferred from the atmospheric oscillation data may be reflected in very large $\tilde{s}_R - \tilde{b}_R$ mixing.
That is, if the neutrinos in generations two and three enjoy large mixing, then quite possibly the right-handed squarks in these two generations do too \cite{r51F}. This  large squark mixing could lead to a large non-Standard-Model contribution to $B_s - \overline{B_s}$ mixing. Revealing the presence of such non-Standard-Model contributions to quark processes is one of the main goals of the present and future exploration of $B$ meson physics.

\section{CONCLUSION}

The evidence that neutrinos change flavor is rich and compelling. This evidence has already taught us something about the neutrino mass spectrum and about the general character of the leptonic mixing matrix. However, there is a lot that we do not know: the absolute scale of neutrino mass, whether neutrinos are their own antiparticles, how large the small mixing angle $\theta_{13}$ is, whether neutrinos violate CP, \ldots, and finally, the physics behind it all. Interesting years lie ahead in neutrino physics.

\section*{ACKNOWLEDGEMENTS}
It is a pleasure to thank the Kavli Institute for Theoretical Physics, where this paper was completed, for its warm hospitality. The author is grateful to Susan Kayser for her crucial role in producing this manuscript.

\def\plb#1#2#3{{\it Phys.\ Lett.} {\bf B#1} (20#2) #3}
\def\plbo#1#2#3{{\it Phys.\ Lett.} {\bf B#1} (19#2) #3}
\def\prd#1#2#3{{\it Phys.\ Rev.} {\bf D#1} (20#2) #3}
\def\prdo#1#2#3{{\it Phys.\ Rev.} {\bf D#1} (19#2) #3}
\def\prl#1#2#3{{\it Phys.\ Rev.\ Lett.} {\bf #1} (20#2) #3}
\def\prlo#1#2#3{{\it Phys.\ Rev.\ Lett.} {\bf #1} (19#2) #3}
\def\etc{{\it et al.}}


\begin{thebibliography}{99}
\bibitem{r7} The leptonic mixing matrix is sometimes referred to as the Maki-Nakagawa-Sakata matrix, or as the Pontecorvo-Maki-Nakagawa-Sakata matrix.
\end{thebibliography}
\end{document}